\documentclass[aps,twocolumn]{revtex4}%
\usepackage{amsfonts}
\usepackage{amsmath}
\usepackage{amssymb}
\usepackage{graphicx}
\usepackage{color}
\usepackage{version}%
\providecommand{\U}[1]{\protect\rule{.1in}{.1in}}

\begin{document}
\title{A fast and robust approach to long-distance quantum communication with atomic ensembles}
\author{L. Jiang$^{1}$, J. M. Taylor$^{1,2}$, M. D. Lukin$^{1}$}
\affiliation{$^{1}$ Department of Physics, Harvard University, Cambridge, Massachusetts 02138}
\affiliation{$^{2}$ Department of Physics, Massachusetts Institute of Technology,
Cambridge, Massachusetts 02139}
\date{\today }

\begin{abstract}
Quantum repeaters create long-distance entanglement between quantum systems
while overcoming difficulties such as the attenuation of single photons in a
fiber. Recently, an implementation of a repeater protocol based on single
qubits in atomic ensembles and linear optics has been proposed [Nature
\textbf{414}, 413 (2001)]. Motivated by rapid experimental progress towards
implementing that protocol, here we develop a more efficient scheme compatible
with active purification of arbitrary errors. Using similar resources as the
earlier protocol, our approach intrinsically purifies leakage out of the
logical subspace and all errors within the logical subspace, leading to
greatly improved performance in the presence of experimental inefficiencies.
Our analysis indicates that our scheme could generate approximately one pair
per 3 minutes over 1280 km distance with fidelity ($F\geq78\%$) sufficient to
violate Bell's inequality.%
\color{black}%

\end{abstract}
\maketitle

\section{Introduction}

Quantum communication holds promise for the secret transfer of classical
messages as well forming an essential element of quantum networks, allowing
for teleportation of arbitrary quantum states and violations of Bell's
inequalities over long distances \cite{GisinRMP}. While experimental and even
commercial implementation of simple quantum communication protocols are well
established \cite{Kwiat98,Zeilinger04}, extending these techniques to
distances much longer than the attenuation length of optical fiber remains a
challenging goal due to exponential attenuation of transmitted signals.
Quantum repeaters \cite{Bennett1996,Deutsch1996,Briegel1998} overcome the
exponential time overhead associated with fiber attenuation and other errors
by using a quantum memory and local quantum computation.

Several promising avenues for quantum repeater implementation include both
atomic ensembles \cite{Duan2001} and using few qubit quantum computers, such
as neutral atoms in cavity QED~\cite{Cirac1997, Kimble2004}, ion
traps~\cite{Blinov2004} and solid-state single photon
emitters~\cite{Childress2004}. Experimental
progress~\cite{Chou2005,Chaneliere2005,Eisaman2005} towards realization of the
DLCZ protocol \cite{Duan2001} has been especially rapid, with many building
blocks demonstrated in the laboratory. The experimental challenge is now
shifting towards the realization of scalable quantum repeater systems which
could yield a reasonable communication rate at continental distances
($\gtrsim1000km$). Thus, the DLCZ protocol should be examined and adapted to
practical experimental considerations, allowing to remove imperfections such
as the finite efficiency of retrieval and single-photon detection and fiber
length fluctuations. Our approach extends the DLCZ protocol, keeping the
experimental simplicity of the original scheme while avoiding fundamental
difficulties due to these expected experimental imperfections.

This paper is organized as follows. In Sec. II we will review the DLCZ
protocol and describe our new approach which uses a new basis to encode each
qubit. Section III compares both the DLCZ protocol and our approach in the
presences of imperfections. Section IV estimates the time scaling of our
approach and compares three specific implementations. Section V summarizes our results.

\section{Atomic-ensemble-based Quantum Repeaters}

\subsection{The DLCZ protocol: a review}

The DLCZ protocol \cite{Duan2001} starts with entanglement generation (ENG) by
counting the interfering Stokes photons scattered from a pair of distant
atomic cells $x$ and $y$. This generates an entangled state
\begin{equation}
\left\vert \xi_{\phi}\right\rangle _{x,y}=\left(  \hat{S}_{x}^{\dag}+e^{i\phi
}\hat{S}_{y}^{\dag}\right)  /\sqrt{2}\left\vert \mathrm{vac}\right\rangle
_{x,y}\ , \label{ENG}%
\end{equation}
with $\hat{S}_{x}^{\dag}$ and $\hat{S}_{y}^{\dag}$ the creation operators of
spin-wave modes in the two cells respectively, and $\phi$ the phase difference
between left and right channels for Stokes photons~\cite{LukinRMP}. Then,
entanglement connection (ENC) is performed on two pairs of entangled atomic
cells $\left\vert \xi_{\phi_{1}}\right\rangle _{x_{L},y_{C}}$ and $\left\vert
\xi_{\phi_{2}}\right\rangle _{x_{C},y_{R}}$, obtaining a further separated
entangled pair $\left\vert \xi_{\phi_{1}+\phi_{2}}\right\rangle _{x_{L},y_{R}%
}$ probabilistically. The ENC step provides built-in purification against many
imperfections -- photon loss, atomic excitation loss and dark counts. In the
final step, post-selection is used to obtain an effectively \emph{polarization
entangled state}
\begin{equation}
\left\vert \Psi^{PME}\right\rangle =e^{i\phi}\left(  \hat{S}_{x_{1}}^{\dag
}\hat{S}_{y_{2}}^{\dag}+\hat{S}_{x_{2}}^{\dag}\hat{S}_{y_{1}}^{\dag}\right)
/\sqrt{2}\left\vert \mathrm{vac}\right\rangle \label{PME}%
\end{equation}
from two parallel pairs $\left\vert \xi_{\phi}\right\rangle _{x_{1},y_{1}%
}\left\vert \xi_{\phi}\right\rangle _{x_{2},y_{2}}$, which overcomes static
phase errors (time independent $\phi$'s).

There are two important merits of the DLCZ protocol. First, it has intrinsic
purification of errors due to photon loss (in the fiber, the quantum memory,
and the photon detector) and significantly relaxes the experimental
requirement for quantum repeater. In addition, the time scaling of the DLCZ
protocol is always sub-exponential and very close to polynomial when the
retrieval and detection efficiency is high. However, the DLCZ protocol does
not purify all kinds of errors. For example, time dependent $\phi$'s (due to
fiber length fluctuation) induce phase error, which cannot be taken out as a
common factor in Eq.(\ref{PME}), since the two pairs of entangled atomic cells
are not produced at the same time. Such phase error is accumulated and doubled
after each level of ENC. In addition, combined photon loss during ENG and ENC
may also induce phase error not purified by the DLCZ protocol. Furthermore,
the DLCZ protocol (dashed line in Fig. 2) still has a significant time
overhead for long distances, because of the super-polynomial scaling in the
presence of realistic imperfections. For instance, non-ideal retrieval and
detection efficiency ($\eta<1$) during ENC introduces a large vacuum
component, suppresses the success probability of later ENC, and consequently
slows down the protocol.

Motivated by these issues, we will extend the DLCZ protocol, mitigating the
above errors.%

\begin{figure}
[ptb]
\begin{center}
\includegraphics[
height=2.2001in,
width=3.0718in
]%
{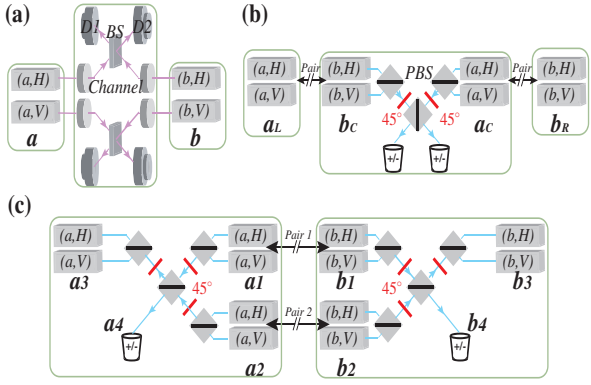}%
\caption[fig1]{Repeater components: (a) Entanglement generation (ENG). (b)
Entanglement connection (ENC); indicated operations: retrieve $b_{C}$ and
$a_{C}$ [additional $45^{\circ}$ rotations only for the first level], join on
polarizing beam splitter (PBS), detect in $\pm$ basis conditioned on one
photon per output, and finally adjust the phase. (c) Entanglement purification
(ENP); indicated operations: retrieve $a_{1},b_{1}$ and $a_{2},b_{2}$
[additional $45^{\circ}$ rotations to purify phase error], interfere
$a_{1},a_{2}$ on PBS (same with $b_{1},b_{2}$), restore $a_{3},b_{3}$
conditioned on single photon at $a_{4}$ and $b_{4}$ respectively, and finally
adjust the phase.}%
\end{center}
\end{figure}

\subsection{New approach}

We now consider a different approach in which two atomic cells are used at
each node $a$, labeled $\left(  a,H\right)  $ and $\left(  a,V\right)  $, to
store one qubit, $a$. The qubit is defined as one single spin-wave excitation
shared between two cells:
\begin{equation}
\left\{  \left\vert H\right\rangle _{a}=S_{a,H}^{\dag}\left\vert
\mathrm{vac}\right\rangle ,\left\vert V\right\rangle _{a}=S_{a,V}^{\dag
}\left\vert \mathrm{vac}\right\rangle \right\}  \ .
\end{equation}
When the stored spin waves are converted back into photons, the photons have a
polarization ($H$ or $V$) consistent with\ that stored in the originating
cell. This qubit basis allows projective measurements along any qubit states,
e.g., $\left\vert \pm\right\rangle \equiv\left(  \left\vert H\right\rangle
_{a}\pm\left\vert V\right\rangle _{a}\right)  /\sqrt{2}$, using linear optical
operations and photon counting \cite{Knill2001}. We will show that in this
logical basis it is possible to perform entanglement purification (ENP)
\cite{Briegel1998} to reduce errors within the logical subspace, including
phase fluctuation. Since ENP can suppress errors within the logical subspace
which occur with probability $q$ to $O\left(  q^{2}\right)  $, only a few ENP
levels are needed to obtain high fidelity entanglement.%

\begin{figure}
[ptb]
\begin{center}
\includegraphics[
height=2.6498in,
width=3.3278in
]%
{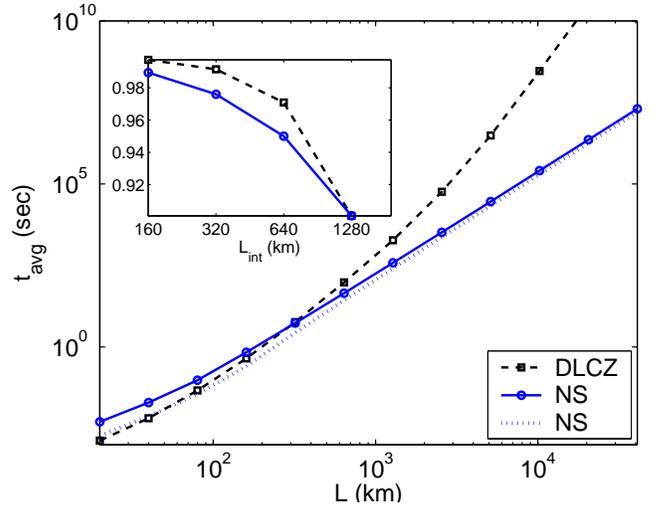}%
\caption[fig2]{Comparison between the DLCZ protocol and the new scheme (NS)
(without active entanglement purification (ENP)). For each distance, we
optimize over the choice of the control parameters (the half distance between
neighboring repeater stations, $L_{0}$, and the elementary pair generation
probability, $p_{c}$). With targeting fidelity $F=90\%$, we find the most
efficient implementations to create the polarization entangled state
(Eq.(\ref{PME})) for both the DLCZ (circled black dashed line) and the new
scheme (squared blue solid line). The fiber attenuation length is
$L_{att}=20km$, with no dynamical phase error. The main plot: we show the
relationship between the (optimized) average creation time $t_{avg}$ and the
final distance $L$ for both schemes, and the empirical estimate
(Eq.(\ref{Empirical})) of the time scaling for the new scheme (blue dotted
line). Over long distances ($L\geq320km$), the polynomial scaling of the new
scheme is more favorable than the super-polynomial scaling of the DLCZ
protocol. The inset: we plot the fidelities of the intermediate distances
($L_{int}=160,320,640$ and $1280km$), to create polarization entangled states
($L=1280$, $F=90\%$), with the optimized choice of the control parameters
$\left(  L_{0},p_{c}\right)  =\left(  80,0.0027\right)  $ and $\left(
40,0.0081\right)  $ for the DLCZ $\left(  t_{avg}\approx1900\sec\right)  $ and
the new scheme $\left(  t_{avg}\approx380\sec\right)  $, respectively. The
optimized choices of the control paramters are detailed in Table
\ref{TableDLCZ} and \ref{TableNS}.}%
\end{center}
\end{figure}

We now describe our procedures for ENG, ENC and ENP. ENG (Fig. 1(a)) is
similar to that of the DLCZ protocol, but here \emph{two} parallel entangled
pairs are generated between $a$ and $b$:
\begin{align}
\left\vert \Psi^{ENG}\right\rangle _{a,b}  &  =\left\vert \xi_{\phi
}\right\rangle _{\left(  a,H\right)  \left(  b,H\right)  }^{+}\left\vert
\xi_{\phi}\right\rangle _{\left(  a,V\right)  \left(  b,V\right)  }%
^{+}\nonumber\\
&  =e^{i\phi}\left(  \left\vert H\right\rangle _{a}\left\vert V\right\rangle
_{b}+\left\vert V\right\rangle _{a}\left\vert H\right\rangle _{b}\right)  +\\
&  \left\vert HV\right\rangle _{a}\left\vert \mathrm{vac}\right\rangle
_{b}+e^{2i\phi}\left\vert \mathrm{vac}\right\rangle _{a}\left\vert
HV\right\rangle _{b}\ .\nonumber
\end{align}
The entangled states are prepared in the quantum memory, so no simultaneity is
required for creating the two states comprising $\left\vert \Psi
^{ENG}\right\rangle $. For small excitation probability $p_{c}$, the whole
generation only takes time $O\left(  1/p_{c}\right)  $, in contrast to
$O\left(  1/p_{c}^{2}\right)  $ for schemes requiring simultaneity, e.g.,
coupling between trapped atom and photon \cite{Blinov2004} or parametric down
conversion \cite{Pan2003}. Errors from multi-photon events occur only with
probability $p_{c}^{2}$, and are considered in later analysis of imperfections.

The first level of ENC converts two $\left\vert \Psi^{ENG}\right\rangle $
states (one between $a_{L}$ and $b_{C}$, the other between $a_{C}$ and $b_{R}
$) into polarization entangled states $\left\vert \Phi^{+}\right\rangle
_{ab}=\left\vert H\right\rangle _{a_{L}}\left\vert H\right\rangle _{b_{R}%
}+\left\vert V\right\rangle _{a_{L}}\left\vert V\right\rangle _{b_{R}} $. Only
four out of the sixteen terms in the Schmidt decomposition of $\left\vert
\Psi^{ENG}\right\rangle _{a_{L}b_{C}}\left\vert \Psi^{ENG}\right\rangle
_{a_{C}b_{R}}$ have any contribution to the output state; the remainder are
eliminated by projective measurement during ENC, reducing the probability of
success for ENC from $1/2$ to $1/8$. At higher levels of ENC, the operations
correspond to standard entanglement swapping \cite{Bennett1996,Deutsch1996},
where
\[
\left\vert \Phi^{\pm}\right\rangle _{a_{L}b_{C}}\otimes\left\vert \Phi^{\pm
}\right\rangle _{a_{C}b_{R}}\rightarrow\left\vert \Phi^{+}\right\rangle
_{a_{L}b_{R}}%
\]
leads to an entangled pair between $L$ and $R$ with probability $1/2$, as
detailed below.

The procedure for ENC is illustrated in Fig.~1(b). First, the spin waves
stored in qubit $b_{C}$ and $a_{C}$ are retrieved into photons. At the lowest
level of ENC, the polarization of the photons is rotated $45^{\circ}$.\ The
rotations transforms $\left\vert HV\right\rangle _{a_{C}\mathrm{(or~}%
b_{C}\mathrm{)}}$\ into $\left(  \left\vert HH\right\rangle -\left\vert
VV\right\rangle \right)  _{a_{C}\mathrm{(or~}b_{C}\mathrm{)}}$, because for
bosonic fields \cite{Merzbacher1998}%

\[
S_{H}^{\dag}S_{V}^{\dag}\overset{45^{\circ}}{\longrightarrow}\left(
S_{H}^{\dag}+S_{V}^{\dag}\right)  \left(  S_{H}^{\dag}-S_{V}^{\dag}\right)
=\left(  S_{H}^{\dag}S_{H}^{\dag}-S_{V}^{\dag}S_{V}^{\dag}\right)  ~.
\]
Thus, after the polarizing beam splitter (PBS) there will be at least two
photons at one output. For incoming state $\left\vert \Psi^{ENG}\right\rangle
_{a_{L}b_{C}}\otimes\left\vert \Psi^{ENG}\right\rangle _{a_{C}b_{R}}$, all
seven terms containing two excitations in at least one pair of cells in the
center repeater node (such as $\left\vert HV\right\rangle _{a_{C}%
\mathrm{(or~}b_{C}\mathrm{)}}$) do not contribute to the click patterns with
one photon at each output. Five terms containing two excitations in one of the
left or right repeater nodes (e.g., $\left\vert HV\right\rangle _{a_{L}%
\mathrm{(or~}b_{R}\mathrm{)}}$) have at most one excitation retrieved from
$b_{C}$\ and $a_{C}$, which is insufficient to give two clicks. Therefore,
only the four terms remaining can give the correct photon detector click patterns.%

\begin{table}[tbp] \centering
\begin{tabular}
[c]{|p{2.5cm}|c|}\hline
\textbf{Operation} & \textbf{Transform of }$\left\vert \Phi^{\pm}\right\rangle
_{a_{L}b_{C}}\left\vert \Phi^{\pm}\right\rangle _{a_{C}b_{R}}$\\\hline\hline
Retrieve $b_{C},a_{C}$ & $\left\vert 0000\right\rangle \pm\left\vert
0011\right\rangle \pm\left\vert 1100\right\rangle +\left\vert
1111\right\rangle $\\\hline
Transform $b_{C},a_{C}$ at PBS &
\begin{tabular}
[c]{l}%
$\left\vert 0000\right\rangle +\left\vert 1111\right\rangle $\\
$~\pm\left\vert 0\right\rangle _{a_{L}}\left(  \left\vert HV\right\rangle
_{a_{C}}\right)  \left\vert 1\right\rangle _{b_{R}}$\\
$~\pm\left\vert 1\right\rangle _{a_{L}}\left(  \left\vert HV\right\rangle
_{b_{C}}\right)  \left\vert 0\right\rangle _{b_{R}}$%
\end{tabular}
\\\hline
One photon per mode ($p=0.5$) & $\left\vert 0000\right\rangle +\left\vert
1111\right\rangle $\\\hline
Detect in $\pm$, results $m,m^{\prime}$ & $\left\vert 00\right\rangle
_{a_{L}b_{R}}+mm^{\prime}\left\vert 11\right\rangle _{a_{L}b_{R}}$\\\hline
Phase shift $mm^{\prime}$ & $\left\vert \Phi^{+}\right\rangle _{a_{L}b_{R}}%
$\\\hline
\end{tabular}
\caption{
Entanglement connection procedure applied to $\left\vert \Phi ^{\pm }\right\rangle $ inputs for entangled pairs between $a_{L}$ and $b_{C}$, and $a_{C}$ and $b_{R}$. For clarity, we introduce $\left\vert 0\right\rangle \equiv \left\vert H\right\rangle $ and $\left\vert
1\right\rangle \equiv \left\vert V\right\rangle $ to represent logical
states (i.e. states with exactly one excitation), $\left\vert
HV\right\rangle $ for non-logical states with two excitations, and $\left\vert \mathrm{vac}\right\rangle $ for states with no excitation
(sometimes omitted).
We assume $H$ photons pass through and $V$ photons be reflected at the middle PBS.
}\label{TableEC}%
\end{table}%

For all levels of ENC, the photons are then joined on the middle PBS and the
number of photons at two outputs are counted in the $\left\{  \left\vert
+\right\rangle ,\left\vert -\right\rangle \right\}  $ basis. With probability
$50\%$, there is one photon at each output, and the connection is successful;
otherwise the process is repeated. If the two photons have orthogonal
polarizations, a bit flip $\alpha\left\vert H\right\rangle +\beta\left\vert
V\right\rangle \rightarrow\alpha\left\vert V\right\rangle +\beta\left\vert
H\right\rangle $ is applied to $a_{L}$~\footnote[4]{The bit and phase flips
called for in ENC and ENP can be performed using linear optics the next time
the qubits are retrieved from the quantum memory.}. At higher levels of ENC,
where the $45^{\circ}$ rotations are not necessary, the bit flip is replaced
by the phase flip $\alpha\left\vert H\right\rangle +\beta\left\vert
V\right\rangle \rightarrow\alpha\left\vert H\right\rangle -\beta\left\vert
V\right\rangle $, as detailed in Table \ref{TableEC}.~\footnote[10]{ENC can be
summarized as $\Theta_{\mathrm{ENC}}^{m,m^{\prime}}\left\vert xy\right\rangle
_{bc}\rightarrow\left(  x\oplus y\oplus1\right)  (mm^{\prime})$ where
$mm^{\prime}$ represents the parity of two detected photons and the logical
states are $\left\vert 0\right\rangle =\left\vert H\right\rangle $ and
$\left\vert 1\right\rangle =\left\vert V\right\rangle $.}

The third component is ENP (Fig.~1(c)) which obtains a high fidelity entangled
pair from two pairs. Our procedure uses polarization entangled photons and is
similar to recent experimental investigations~\cite{Pan2003}. During
entanglement purification of bit errors (bit-ENP), the qubits from two
parallel pairs $\rho_{a_{1},b_{1}}$ and $\rho_{a_{2},b_{2}}$ are retrieved
from the quantum memory and joined at PBSs. The photons for two upper outputs
are stored into quantum memory $a_{3}$ and $b_{3}$. The photons for the lower
outputs $a_{4}$ and $b_{4}$ are counted in $\left\{  \left\vert +\right\rangle
,\left\vert -\right\rangle \right\}  $ basis. With probability $50\%$, there
is exactly one photon at each lower output, and the purification is
successful; otherwise two new pairs are created by restarting the process. If
the two photons have orthogonal polarizations, a phase flip is applied to
$a_{3}$. An example of purification of bit-error is presented in Table
\ref{TableEP}. During purification of phase errors (phase-ENP), additional
$45^{\circ}$ rotations are applied to the retrieved qubits and the bit flip is
replaced by the phase flip. The addition of $45^{\circ}$ rotations effects the
basis transform $\left\vert \Phi^{-}\right\rangle \leftrightarrow\left\vert
\Psi^{+}\right\rangle $, leading to purification of errors of the other type.
The truth table of phase-ENP is listed in Table \ref{TableEP2}. Bit (or phase)
errors can be non-linearly suppressed to the second order during bit-ENP (or
phase-ENP) \footnote[15]{ENP can be summarized as $\Theta_{ENP}^{m,m^{\prime}%
}\left\vert x\right\rangle _{a_{1}}\left\vert y\right\rangle _{b_{1}%
}\left\vert u\right\rangle _{a_{2}}\left\vert v\right\rangle _{b_{2}%
}\rightarrow\left(  x\oplus u\oplus1\right)  \left(  y\oplus v\oplus1\right)
\left(  -1\right)  ^{x\cdot\left(  mm^{\prime}+1\right)  }\left\vert
x\right\rangle _{a_{4}}\left\vert y\right\rangle _{b_{4}}$ with binary basis
$\left\{  \left\vert 0\right\rangle ,\left\vert 1\right\rangle \right\}  $ for
$\left\{  \left\vert H\right\rangle ,\left\vert V\right\rangle \right\}  $
during bit-ENP and $\left\{  \left\vert +\right\rangle ,\left\vert
-\right\rangle \right\}  $ during phase-ENP, and $mm^{\prime}$ represents the
parity of two detected photons.}.%

\begin{table}[tbp] \centering
\begin{tabular}
[c]{|p{2.5cm}|c|}\hline
\textbf{Operation} & \textbf{Transform of }$\left\vert \Phi^{n}\right\rangle
_{a_{1}b_{1}}\left\vert \Psi^{n^{\prime}}\right\rangle _{a_{2}b_{2}}%
$\\\hline\hline
Retrieve $a_{1}b_{1},a_{2}b_{2}$ & $\left\vert 0001\right\rangle +n\left\vert
1101\right\rangle +n^{\prime}\left\vert 0010\right\rangle +nn^{\prime
}\left\vert 1110\right\rangle $\\\hline
Interfere $a_{1},a_{2}$ on PBS; same with $b_{1},b_{2}$ &
\begin{tabular}
[c]{l}%
$\left\vert 00\right\rangle _{a_{3}a_{4}}\left(  \left\vert HV\right\rangle
_{b_{4}}\right)  $\\
$~+n\left(  \left\vert HV\right\rangle _{a_{3}}\right)  \left\vert
11\right\rangle _{b_{3}b_{4}}$\\
$~+n^{\prime}\left(  \left\vert HV\right\rangle _{a_{4}}\right)  \left\vert
00\right\rangle _{b_{3}b_{4}}$\\
$~+nn^{\prime}\left\vert 11\right\rangle _{a_{3}a_{4}}\left(  \left\vert
HV\right\rangle _{b_{3}}\right)  $%
\end{tabular}
\\\hline
One photon per lower output & $\mathbf{0}$\\\hline
\end{tabular}
\caption{
Bit entanglement purification acting on $\left\vert
\Phi ^{+}\right\rangle \left\vert \Psi ^{+}\right\rangle $. The same notation as
Table \ref{TableEC} is used. For such given input states, the click pattern
never matches the right pattern, and bit-error of single input qubit can be
filtered completely. The remaining steps to preserve the desired Bell states
are detection of $a_{4},b_{4}$ in $\pm $ basis, a phase shift $n\cdot
n^{\prime }$ based on the measurement results, and storing of the travelling
photons ($a_{3},b_{3}$) into atomic ensembles.}\label{TableEP}%
\end{table}%

The three components described above for quantum repeater protocol only use
atomic cells, linear optics, and photon number counting. We remark that the
duration of the retrieved anti-Stokes pulse can be made long ($\gtrsim1\mu s$)
compared to the detector recovery time by adjusting the intensity and duration
of the retrieval pulse. This enables photon number counting of the anti-Stokes
pulse \cite{Eisaman2004}.%

\begin{table}[tbp] \centering
$%
\begin{tabular}
[c]{|c|c|c|c|c|}\hline
$\rho_{a_{1},b_{1}}~\backslash~\rho_{a_{2},b_{2}}$ & $\Phi^{+}$ & $\Phi^{-}$ &
$\Psi^{+}$ & $\Psi^{-}$\\\hline
$\Phi^{+}$ & $\Phi^{+}$ & $-$ & $\Psi^{+}$ & $-$\\\hline
$\Phi^{-}$ & $-$ & $\Phi^{-}$ & $-$ & $\Psi^{-}$\\\hline
$\Psi^{+}$ & $\Psi^{+}$ & $-$ & $\Phi^{+}$ & $-$\\\hline
$\Psi^{-}$ & $-$ & $\Psi^{-}$ & $-$ & $\Phi^{-}$\\\hline
\end{tabular}
$\caption{
Truth table for phase-ENP. Each element give the
possible output state after the purification operation. ("$-$" for cases
with no outputs.)}\label{TableEP2}%
\end{table}%

\section{Noise and Imperfections}

\subsection{Non-logical errors}

We now examine the performance of our new scheme by considering the role of
errors, starting with how imperfections due to inefficiency limit the
protocols. Primarily, we find that inefficiency takes logical states into two
types of non-logical states -- those with too few excitations (\emph{vacuum}
type) and those with too many excitations (\emph{multi-excitation} type). We
represent these errors by density matrix $\pi_{vac}$ (a mixed state with at
most one excitation between both pairs of cells) or $\pi_{multi}$ (a mixed
state with at least one pair of cells with more than one excitation). The
normalized density matrix after $m$th ENC (Fig. 1(b)) can be written as%
\begin{equation}
\rho_{a_{L},b_{R}}^{\left(  m\right)  }=p_{\mathrm{\emph{logic}}}^{\left(
m\right)  }\rho_{\mathrm{\emph{logic}}}^{\left(  m\right)  }+p_{vac}^{\left(
m\right)  }\pi_{vac}^{\left(  m\right)  }+p_{multi}^{\left(  m\right)  }%
\pi_{multi}^{\left(  m\right)  } \label{rou_m}%
\end{equation}
where the $m$-dependent operator $\rho_{\mathrm{\emph{logic}}}^{\left(
m\right)  }$ is the density matrix within logical subspace; $\pi
_{vac}^{\left(  m\right)  }$ and $\pi_{multi}^{\left(  m\right)  }$ also
depend on $m$; and the coefficients $p_{\mathrm{\emph{logic}}}^{\left(
m\right)  }$, $p_{vac}^{\left(  m\right)  }$ and $p_{multi}^{\left(  m\right)
}$ are the probabilities for the logical, vacuum and multi-excitation types, respectively.

After the first level of ENC, $p_{vac}^{\left(  1\right)  }\sim1-\eta$ and
$p_{multi}^{\left(  1\right)  }\sim p_{c}\ll1$. We can demonstrate that these
three probabilities remain stable (see Appendix B) for all higher levels of
ENC, by considering the un-normalized state after $\left(  m+1\right)  $th ENC
for the new approach: $\tilde{\rho}_{a,b}^{\left(  m+1\right)  }=\tilde
{p}_{\mathrm{\emph{logic}}}^{\left(  m+1\right)  }\rho_{\mathrm{\emph{logic}}%
}^{\left(  m+1\right)  }+\tilde{p}_{vac}^{\left(  m+1\right)  }\pi
_{vac}^{\left(  m+1\right)  }+\tilde{p}_{multi}^{\left(  m+1\right)  }%
\pi_{multi}^{\left(  m+1\right)  }$, with%
\begin{align*}
\tilde{p}_{\mathrm{\emph{logic}}}^{\left(  m+1\right)  }  &  \approx\frac
{1}{2}\eta p_{\mathrm{\emph{logic}}}^{\left(  m\right)  }%
p_{\mathrm{\emph{logic}}}^{\left(  m\right)  }\left(  1+p_{err,new}^{\left(
m+1\right)  }+O\left(  p_{c}\right)  \right) \\
\tilde{p}_{vac}^{\left(  m+1\right)  }  &  \approx\frac{1}{2}\eta
p_{\mathrm{\emph{logic}}}^{\left(  m\right)  }p_{vac}^{\left(  m\right)
}\left(  1+O\left(  p_{c}\right)  \right) \\
\tilde{p}_{multi}^{\left(  m+1\right)  }  &  \approx\frac{1}{2}\eta
p_{\mathrm{\emph{logic}}}^{\left(  m\right)  }p_{multi}^{\left(  m\right)
}\left(  1+O\left(  p_{c}\right)  \right)
\end{align*}
where the probability for the new logical error from the multi-excitation
states (accompanied by photon loss) is%
\[
p_{err,new}^{\left(  m+1\right)  }\sim\left(  1-\eta\right)  p_{c}.
\]
The total logical error probability $p_{err}^{\left(  m+1\right)  }$ has two
contributions: the accumulated logical errors from both input pairs for ENC,
$p_{err}^{\left(  m\right)  }$, and the new logical error%
\begin{equation}
p_{err}^{\left(  m+1\right)  }-2p_{err}^{\left(  m\right)  }\sim\left(
1-\eta\right)  p_{c}. \label{LogicalError}%
\end{equation}
We can calculate for the new scheme,%
\begin{equation}
p_{err}^{\left(  m\right)  }\sim\left(  2^{m}-1\right)  \left(  1-\eta\right)
p_{c}. \label{LogicalError2}%
\end{equation}

A more detailed calculation (see Appendix B) , in which $\pi_{vac}$ and
$\pi_{multi}$ are further divided into subspaces with different number of
excitations (e.g. $\pi_{vac}$ is subdivided into zero-excitation and
one-excitation subspaces), verifies the stability of the probability
distribution of $p_{\emph{logic}}^{\left(  m\right)  }$, $p_{vac}^{\left(
m\right)  }$ and $p_{multi}^{\left(  m\right)  }$. (Similarly, dark count can
also induce errors in logical subspace with probability $\sim p_{dark}\left(
1-\eta_{s}\right)  $, which is however negligible due to very low dark count
probability $p_{dark}$.)

For the DLCZ protocol, only two cells are used to store entanglement. Besides
the logical states (single excitation in two cells), we can similarly define
the vacuum states (with no excitation) and multi-excitation states (with two
or more excitations). Contrary to our approach, the probability distribution
is not stable -- both vacuum and multi-excitation probabilities increases with
distance (see Appendix A). The vacuum probability soon becomes the dominant
term, which reduces the success probability of ENC significantly, resulting in
super-polynomial (but still sub-exponential) time scaling (Fig. 2(b)). The
logical error probability for the DLCZ protocol has the same form as
Eq.(\ref{LogicalError}) up to the coefficient, but the ratio $p_{multi}%
^{\left(  m\right)  }/p_{\emph{logic}}^{\left(  m\right)  }$ (thus
$p_{err}^{\left(  m\right)  }$) grows with distance (see Appendix A), which
accounts for the sharp decrease of fidelity for the DLCZ protocol in Fig.
2(a). To maintain good final fidelity, the initial error $p_{multi}^{\left(
1\right)  }$ (and $p_{c}$) should be very small, which demands much longer
generation time of an elementary pair for the DLCZ protocol (Fig. 2(b)).

In essence, by requiring at least one excitation in the ensemble, our qubit
subspace is automatically purified of vacuum and multi-excitation type errors
during ENC. The closest analog to the DLCZ protocol is our new scheme without
ENP, i.e., only ENC. At longer distances, our approach is further improved in
comparison to the DLCZ protocol due to the reduced amplitude of vacuum terms.

\subsection{Logical Errors}

So far, we have only considered the effects of inefficiency that maps states
between logical and non-logical subspace by changing the number of
excitations. Besides inefficiency, there are other imperfections, which
preserve the number of excitations but induce errors within the logical
subspace, such as interferometric pathlength fluctuation and linear optical
misalignment. For example, the interferometric pathlength fluctuation leads to
a phase difference $\delta$, which changes $\left\vert \Psi^{ENG}\right\rangle
_{a,b}$ into a mixture:
\begin{align}
\left\vert \Psi^{ENG}\right\rangle _{a,b}  &  \rightarrow e^{i\phi}\left(
\left\vert H\right\rangle _{a}\left\vert V\right\rangle _{b}+e^{i\delta
}\left\vert V\right\rangle _{a}\left\vert H\right\rangle _{b}\right) \\
&  +\left\vert HV\right\rangle _{a}\left\vert \mathrm{vac}\right\rangle
_{b}+e^{2i\phi+i\delta}\left\vert \mathrm{vac}\right\rangle _{a}\left\vert
HV\right\rangle _{b}\ .\nonumber
\end{align}
where $\phi$ is static phase difference between left and right channels. Since
the last two terms with $\left\vert HV\right\rangle $ will be removed during
the first level of ENC, the static phase $\phi$ has no effect. However, the
probability of being in an undesired logical state $\left\vert \Psi
^{-}\right\rangle $ is $\sin^{2}\frac{\delta}{2}$. The first level of ENC with
the combined two inputs of $\Psi^{-}$ and $\Psi^{+}$ gives $\Phi^{-} $,
producing a phase error with probability $2p_{phase-err}=2\left\langle
\sin^{2}\frac{\delta}{2}\right\rangle $, proportional to the variance of the
interferometric phase fluctuation. This error will be amplified during
subsequent ENC's, because the survival probability of the state $\Phi^{-}$
(the logical error) is twice as much as that of $\Phi^{+}$ (the desired component).

\label{sec:DiffusionModel}We expect that there is little correlation in phase
fluctuation between different sections of the fiber, and the variance of the
phase fluctuation is proportional to the length of the fiber%
\begin{equation}
\left\langle \delta^{2}\right\rangle =2DL_{0},
\end{equation}
where $D$ is the phase diffusion coefficient of the fiber. If the phase
fluctuation satisfies gaussian distribution, the phase error probability
\begin{equation}
p_{phase-err}=\left\langle \sin^{2}\frac{\delta}{2}\right\rangle =\frac{1}%
{2}\left(  1-e^{-DL_{0}}\right)  .
\end{equation}
For example, $D=10^{-3}rad^{2}/km$, $L_{0}=10km$, and $p_{phase-err}%
\approx0.5\%$.

Also, a small probability ($p_{err}$) of linear optical misalignment per ENC
or ENP step is modeled as depolarizing errors. Later, we will demonstrate that
errors within the logical subspace restrict the final fidelity of the DLCZ
protocol, while for our new approach additional active purification can
correct such logical errors to achieve high fidelity.

\section{Scaling and Time Overhead for Quantum Repeater}

\subsection{Scaling analysis}

Based on the calculation of the success probability at each level of
connection/purification, we can obtain the estimated average time for various
schemes. In Fig. 2, we compare our approach to the DLCZ protocol. For the DLCZ
protocol, the average creation time for a distant pair contains a
super-polynomial contribution (but still sub-exponentially) with distance, due
to instability of the vacuum component. For our new scheme, the scaling is
strictly polynomial with distance, $t_{avg}\propto L^{\alpha}$, where the
exponent $\alpha=\alpha\left(  \eta\right)  $ explicitly depends on the efficiency.

We remark that the DLCZ protocol is slightly more efficient for short final
distances ($L\leq160km$). The DLCZ protocol can skip the entanglement
connection and exploits post-selection to create the polarization entangled
state, while the new scheme requires the first level of ENC to eliminate those
unwanted components and prepare the polarization entangled state. The
post-selection has success probability $1/2$, two times more efficient than
the success probability $1/4$ for the first level of ENC of the new scheme.

For the new scheme (without ENP), we can use the stable probability
distribution to estimate the average time:%

\begin{equation}
t_{avg}=t_{0}\left(  L/L_{0}\right)  ^{\log_{2}\left(  1.5~\frac{2\left(
2-\eta\right)  ^{4}}{\eta^{2}\left(  3-2\eta\right)  }\right)  }
\label{Empirical}%
\end{equation}
where $t_{0}=\frac{1}{p_{c}\eta}\frac{L_{0}}{c}e^{L_{0}/L_{att}}$ is the
elementary pair generation time, $p_{C}$ is the elementary pair generation
probability, $L_{0}$ is half the distance between neighboring repeater
stations, and $L$ is the final distance. The exponent can be understood as an
overhead from the finite success probability for ENC, $p_{ENC}\approx
\frac{\eta^{2}\left(  3-2\eta\right)  }{2\left(  2-\eta\right)  ^{4}}$. The
constant $1.5$ in Eq.(\ref{Empirical}) is the empirical estimate of the
overhead from the waiting time to obtain two independent pairs versus the
single pair. In Fig. 2, the differences between the simulated data and
Eq.(\ref{Empirical}) are attributed to the empirical factor ($1.5$) and an
overall factor from the overestimate of the success probability for the first
level of ENC. According to Eq.(\ref{LogicalError2}), one can always reach good
final fidelity if the elementary pair generation probability $p_{c}$ scales as
$L_{0}/L$. Therefore, the average distant pair generation time scales
\emph{exactly polynomially} with distance $t_{avg}\propto L^{\alpha}$, with
$\alpha=1+\log_{2}\left(  1.5\right)  +\log_{2}\left(  \frac{2\left(
2-\eta\right)  ^{4}}{\eta^{2}\left(  3-2\eta\right)  }\right)  $.%

\begin{figure}
[ptb]
\begin{center}
\includegraphics[
height=2.4016in,
width=3.0381in
]%
{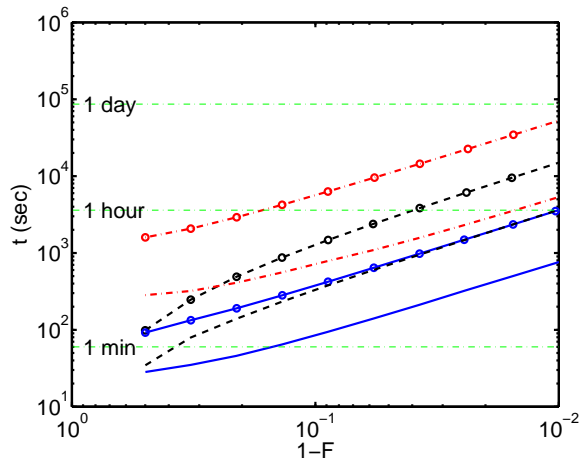}%
\caption[fig3]{Average time versus final fidelity with no phase errors. For
fixed final distance $L=1280$, we plot the (final) fidelity dependence of the
(optimized) average pair creation time, for the DLCZ protocol (black dashed
lines), new scheme without ENP (blue solid lines), and new scheme with ENP
(red dashdotted lines), with efficiency $\eta$ to be $90\%$ (circles), and
$95\%$ (no circles).}%
\end{center}
\end{figure}

\subsection{Comparison between different schemes}

Besides the DLCZ protocol and our new approach without ENP, we now consider a
scheme with ENP, which has one phase-ENP after the second level of ENC. We may
compare these schemes by using $t-F$ plots -- a parametric plot of $t_{avg}$
and $F$ as a function of excitation probability $p_{c}$. For given noise model
and efficiency $\eta$, a repeater scheme corresponds to a curve on $t-F$ plane.

In the absence of interferometric pathlength fluctuation (Fig. 3), the new
approach without ENP is about $5$ times faster than the DLCZ protocol, for
$\eta=90\%$. As given by the previous discussion, this improvement is due to
better control of inefficiency-induced imperfections. There is a time overhead
for the new approach (with ENP) as compared to the new approach (without ENP).
Within each implementation, the higher the efficiency $\eta$, the faster the
quantum repeater. For high final fidelity ($1-F\leq10\%$), the curves approach
straight lines with slope $-1$, because $t\propto p_{c}^{-1}\propto\left(
1-F\right)  ^{-1}$.

When interferometric pathlength fluctuation (leading to initial phase error)
is non-negligible, active ENP is needed. We use a diffusion model for the
pathlength fluctuation, as detailed in Sec.~\ref{sec:DiffusionModel}. In Fig.
4, $t-F$ curves are plotted, assuming the phase diffusion coefficient
$D=10^{-3}rad^{2}/km$, corresponding to $p_{phase-err}\approx0.5\%$ over
$L_{0}=10km$. Unlike Fig. 3 where only inefficiency is considered, there is an
upper bound in final fidelity for each implementation. Both the DLCZ protocol
and new approach (without ENP) suffer from the initial phase error, with final
fidelity no more than $65\%$, while new approach with ENP maintains high final
fidelity up to $97\%$. For high retrieval and detection efficiency
($\eta=95\%$), new approach (with ENP) can produce $1280km$ entangled pairs
with fidelity $90\%$ at a rate of $2$ pairs/hour, even in the presence of
substantial dynamical phase errors.%

\begin{figure}
[ptb]
\begin{center}
\includegraphics[
height=2.4025in,
width=3.0381in
]%
{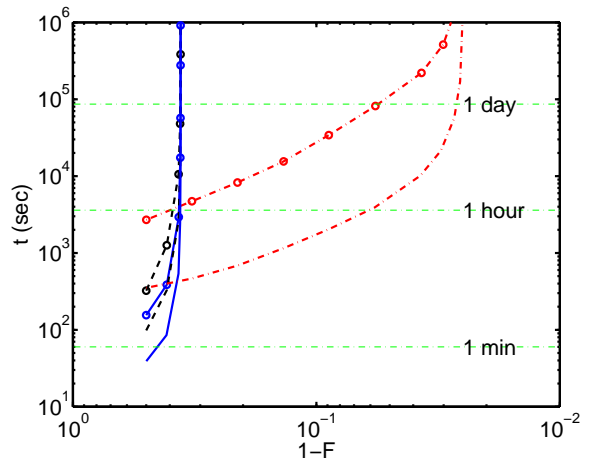}%
\caption[fig4]{Average time versus final fidelity with phase errors. This
shows the same optimized approaches as Fig. 3 but with finite interferometric
pathlength fluctuation characterized by the phase diffusion coefficient
$D=10^{-3}rad^{2}/km$. Inclusion of ENP (red dashdotted lines) yields a
dramatic improvement in time for high fidelity operations.}%
\end{center}
\end{figure}

\section{Outlook}

In summary, our new approach to long distance quantum communication uses a
different qubit basis which prevents the growth of vacuum and multi-excitation
probabilities. This keeps the ENC success probability high and error
probability low, and leads to true polynomial scaling even in the presence of
realistic inefficiencies. We can achieve a bandwidth of $1$ (or $2$) entangled
pair(s) per six minutes for $\eta\approx90\%$ (or $78\%$) and negligible
initial phase error. The new approach also allows active entanglement
purification, which combined with built-in purification of transmission loss
errors allows purification of arbitrary errors in quantum communication.

Although the present approach shows a dramatic improvement in communication
rates and robustness compared to original the DLCZ protocol, the bandwidth
remains relatively slow, even when very high efficiencies and very long-lived
quantum memory are assumed. While such high efficiencies might ultimately be
achievable (see Ref. \cite{Thompson2003} for recent progress), other
approaches need to be considered that can further improve the effective
communication bandwidth. For example, we can use many cells per node to
improve the bandwidth. In this case, the improvement is at least linear with
the number of cells, making it possible to realize long distance ($1280km$)
entangled state generation bandwidth of the order of one pair per second.

A simple Monte Carlo optimization of efficient use of the cells shows that we
can increase the bandwidth by a factor of $r=M^{1.12}$, where $M$ is the
increment of factor of physical resources. Recently, it has also been
suggested that multiple cells can be used to further facilliate the quantum
repeater, in the presence of memory errors from the quantum memory
\cite{Collins06}.

The work was supported by ARO/ARDA, DARPA, NSF Career award, Alfred P. Sloan
Foundation, and David and Lucile Packard Foundation.

\emph{Note added.} -- Since completion of this work, a preprint by Chen
\emph{et al}. \cite{Chen06} describing a similar approach to improve the DLCZ
protocol has appeared.%

\begin{widetext}%
%

\begin{table}[tbp] \centering
\begin{tabular}
[c]%
{c||p{1.5cm}p{1.5cm}p{1.5cm}p{1.5cm}p{1.5cm}p{1.5cm}p{1.5cm}p{1.5cm}p{1.5cm}p{1.5cm}}%
\hline
$L\left(  km\right)  $ & $20$ & $40$ & $80$ & $160$ & $320$ & $640$ & $1280$ &
$2560$ & $5120$ & $10240$\\\hline
$t_{avg}\left(  \sec\right)  $ & $0.0013$ & $0.0065$ & $0.046$ & $0.44$ &
$5.8$ & $95$ & $1900$ & $5.7\times10^{4}$ & $3.0\times10^{6}$ & $3.0\times
10^{8}$\\\hline
$L_{0}\left(  km\right)  $ & $10$ & $20$ & $40$ & $40$ & $40$ & $80$ & $80$ &
$80$ & $80$ & $80$\\\hline
$p_{c}$ & $0.26$ & $0.18$ & $0.13$ & $0.040$ & $0.011$ & $0.010$ &
$2.7\times10^{-3}$ & $7.1\times10^{-4}$ & $1.8\times10^{-4}$ & $4.6\times
10^{-5}$\\\hline
$F_{fin}$ & $90\%\qquad$ & $90\%$ & $90\%$ & $90\%$ & $90\%$ & $90\%$ & $90\%
$ & $90\%$ & $90\%$ & $90\%$\\\hline
\end{tabular}
\caption{Miminized average time $t_{avg}$ and optimized control parameters
($L_{0}$ , $p_c$)  for the DLCZ protocol. This table provides detail
information for Fig. 2.}\label{TableDLCZ}%
\end{table}%
%

\begin{table}[tbp] \centering
\begin{tabular}
[c]%
{l||p{1.5cm}p{1.5cm}p{1.5cm}p{1.5cm}p{1.5cm}p{1.5cm}p{1.5cm}p{1.5cm}p{1.5cm}p{1.5cm}}%
\hline
$L\left(  km\right)  $ & $20$ & $40$ & $80$ & $160$ & $320$ & $640$ & $1280$ &
$2560$ & $5120$ & $10240$\\\hline
$t_{avg}\left(  \sec\right)  $ & $0.0051$ & $0.020$ & $0.10$ & $0.68$ & $5.4$
& $45$ & $380$ & $3.3\times10^{3}$ & $2.9\times10^{4}$ & $2.6\times10^{5}%
$\\\hline
$L_{0}\left(  km\right)  $ & $5$ & $10$ & $20$ & $40$ & $40$ & $40$ & $40$ &
$40$ & $40$ & $40$\\\hline
$p_{c}$ & $0.26$ & $0.17$ & $0.11$ & $0.087$ & $0.037$ & $0.017$ &
$8.1\times10^{-3}$ & $4.0\times10^{-3}$ & $2.0\times10^{-3}$ & $9.7\times
10^{-4}$\\\hline
$F_{fin}$ & $90\%\qquad$ & $90\%$ & $90\%$ & $90\%$ & $90\%$ & $90\%$ & $90\%
$ & $90\%$ & $90\%$ & $90\%$\\\hline
\end{tabular}
\caption{Miminized average time $t_{avg}$ and optimized control parameters
($L_{0}$ , $p_c$)  for the new scheme. This table provides detail
information for Fig. 2.}\label{TableNS}%
\end{table}%

%

\end{widetext}%

\appendix

\section*{Appendix}

We now justify two main claims used in the previous discussion. (1) For the
DLCZ protocol, the probability ratios $p_{vac}^{\left(  m\right)
}/p_{\mathrm{\emph{logic}}}^{\left(  m\right)  }$ and $p_{multi}^{\left(
m\right)  }/p_{\mathrm{\emph{logic}}}^{\left(  m\right)  }$ increase with the
nesting level $m$. (2) For the new scheme (without ENP), the probability
ratios $p_{vac}^{\left(  m\right)  }/p_{\mathrm{\emph{logic}}}^{\left(
m\right)  }$ and $p_{multi}^{\left(  m\right)  }/p_{\mathrm{\emph{logic}}%
}^{\left(  m\right)  }$ remain almost independent of the nesting level $m$.
The states of logical, vacuum and multi-excitation types depend on the
repeater protocol. For example, the entangled logical state for the DLCZ
protocol has total of one excitation stored in two remotely entangled cells
(e.g. Eq.(\ref{ENG})), while the entangled logical state for the new scheme
has two excitations stored in four cells, with one excitation and two cells on
each side (e.g. Eq.(\ref{PME})). Thus, the definitions of
$p_{\mathrm{\emph{logic}}}^{\left(  m\right)  }$, $p_{vac}^{\left(  m\right)
}$ and $p_{multi}^{\left(  m\right)  }$ are different for the two schemes.

\section{Non-logical states for the DLCZ protocol}

We start with the DLCZ protocol. First, we decompose the density matrix (for a
pair of distant atomic cells $x$ and $y$) into components with different
excitation patterns, neglecting the inter-pattern coherence%
\begin{equation}
\rho_{x,y}=p_{00}\pi_{00}+p_{10}\pi_{10}+p_{11}\pi_{11}+p_{20}\pi_{20}+\cdots
\end{equation}
where $p_{ij}\pi_{ij}$ is the projected density matrix to the subspace spanned
by the Fock states of the cells $\left\{  \left\vert i\right\rangle
_{x}\left\vert j\right\rangle _{y},\left\vert j\right\rangle _{x}\left\vert
i\right\rangle _{y}\right\}  $, with probability $p_{ij}$ and normalized
density matrix $\pi_{ij}$. For the DLCZ protocol, the vacuum type of state is
$\pi_{00}$, the logical type of state is $\pi_{10}$, and the rest belong to
the multi-excitation type. We may also introduce the notation corresponding to
Eq.(\ref{rou_m})%
\begin{equation}
\rho_{x,y}=p_{\mathrm{\emph{logic}}}\rho_{\mathrm{\emph{logic}}}+p_{vac}%
\pi_{vac}+p_{multi}\pi_{multi}%
\end{equation}
with $p_{vac}=p_{00}$, $p_{\mathrm{\emph{logic}}}=p_{10}$, and $p_{multi}%
=p_{11}+p_{20}+\cdots$.

Since we are only interested in the coherence properties for the logical type
$\pi_{10}$, we only keep track of the probabilities for the vacuum and
multi-excitation types and neglect their coherences. From the symmetry for the
two cells, we have
\begin{equation}
\pi_{ij}=\pi_{ji}\equiv\frac{1}{2}\left(  \left\vert i\right\rangle
_{x}\left\langle i\right\vert \otimes\left\vert j\right\rangle _{y}%
\left\langle j\right\vert +\left\vert j\right\rangle _{x}\left\langle
j\right\vert \otimes\left\vert i\right\rangle _{y}\left\langle i\right\vert
\right)  .
\end{equation}
In the rest of this subsection, we will add a superscript to $\pi_{ij}$ only
when we want to keep track of the coherence for that specific term. For
example, $\pi_{10}^{\left(  m\right)  }$ indicates that there is coherence
between the states $\left\vert 1\right\rangle _{x}\left\vert 0\right\rangle
_{y}$ and $\left\vert 0\right\rangle _{x}\left\vert 1\right\rangle _{y}$,
after the $m$th level of ENC.

Since the DLCZ protocol requires that the probability for the multi-excitation
states should always be much smaller than the probability for the logical
states (otherwise, a large fraction of multi-excitation states, accompanied by
photon loss, can induce significant logical errors), we regard $p_{11}%
^{\left(  m\right)  }$, $p_{20}^{\left(  m\right)  }$($\ll p_{10}^{\left(
m\right)  }$) as perturbations.

We denote the entangled state after the $m$th ENC as%
\begin{equation}
\rho_{x,y}^{\left(  m\right)  }=p_{00}^{\left(  m\right)  }\pi_{00}%
+p_{10}^{\left(  m\right)  }\pi_{10}^{\left(  m\right)  }+p_{11}^{\left(
m\right)  }\pi_{11}+p_{20}^{\left(  m\right)  }\pi_{20}+\cdots
\end{equation}
We now connect two such entangled states $\rho_{x_{L},y_{L}}^{\left(
m\right)  }$ and $\rho_{x_{R},y_{R}}^{\left(  m\right)  }$ for the $\left(
m+1\right)  $th ENC via the superoperator $\mathcal{E}_{ENC}$
\begin{equation}
\tilde{\rho}_{x_{L},y_{R}}^{\left(  m+1\right)  }=\mathcal{E}_{ENC}\left[
\rho_{x_{L},y_{L}}^{\left(  m\right)  },\rho_{x_{R},y_{R}}^{\left(  m\right)
}\right]  .
\end{equation}
where $\tilde{\rho}_{x_{L},y_{R}}^{\left(  m+1\right)  }$ is the unnormalized
density matrix for the entangled state after the $\left(  m+1\right)  $th ENC.
Since $\mathcal{E}_{ENC}$ is a linear operator and two inputs have the same
state,
\begin{equation}
\mathcal{E}_{ENC}\left[  \sum_{\alpha}p_{\alpha}\pi_{\alpha},\sum_{\beta
}p_{\beta}\pi_{\beta}\right]  =\sum_{\alpha,\beta}p_{\alpha}p_{\beta
}\mathcal{E}_{ENC}^{sym}\left[  \pi_{\alpha},\pi_{\beta}\right]  ,
\end{equation}
where $\mathcal{E}_{ENC}^{sym}\left[  \pi_{\alpha},\pi_{\beta}\right]
\equiv\frac{1}{2}\mathcal{E}_{ENC}\left[  \pi_{\alpha},\pi_{\beta}\right]
+\frac{1}{2}\mathcal{E}_{ENC}\left[  \pi_{\beta},\pi_{\alpha}\right]  $. Now
we calculate $\mathcal{E}_{ENC}^{sym}\left[  \pi_{\alpha},\pi_{\beta}\right]
$ for $\pi_{\alpha},\pi_{\beta}\in\left\{  \pi_{00},\pi_{10}^{\left(
m\right)  },\pi_{11},\pi_{20},\cdots\right\}  $. For example,%
\begin{subequations}
\begin{align}
\mathcal{E}_{ENC}^{sym}\left[  \pi_{10}^{\left(  m\right)  },\pi_{10}^{\left(
m\right)  }\right]   &  =\frac{\eta}{2}\pi_{10}^{\left(  m+1\right)  \prime
}+\frac{\eta\left(  1-\eta\right)  }{2}\pi_{00}\\
\mathcal{E}_{ENC}^{sym}\left[  \pi_{10}^{\left(  m\right)  },\pi_{00}\right]
&  =\frac{\eta}{2}\pi_{00}\\
\mathcal{E}_{ENC}^{sym}\left[  \pi_{10}^{\left(  m\right)  },\pi_{11}\right]
&  =\frac{\eta}{2}\pi_{11}+\eta\left(  1-\eta\right)  \pi_{10}\\
\mathcal{E}_{ENC}^{sym}\left[  \pi_{10}^{\left(  m\right)  },\pi_{20}\right]
&  =\frac{\eta}{4}\pi_{20}+\frac{\eta\left(  1-\eta\right)  }{2}\pi_{10}\\
&  +\frac{3\eta\left(  1-\eta\right)  ^{2}}{4}\pi_{00}\\
\mathcal{E}_{ENC}^{sym}\left[  \pi_{00},\pi_{00}\right]   &  =0\\
\mathcal{E}_{ENC}^{sym}\left[  \pi_{00},\pi_{11}\right]   &  =\eta\pi_{10}\\
\mathcal{E}_{ENC}^{sym}\left[  \pi_{0},\pi_{20}\right]   &  =\eta\left(
1-\eta\right)  \pi_{00}.
\end{align}
The logical type of state after the $\left(  m+1\right)  $th ENC, $\pi
_{10}^{\left(  m+1\right)  }$, is the average of the states $\pi_{10}^{\left(
m+1\right)  \prime}$ and $\pi_{10}$, with relative weights $\frac{\eta}%
{2}p_{10}^{\left(  m\right)  }p_{10}^{\left(  m\right)  }$ and $2\eta\left(
1-\eta\right)  p_{10}^{\left(  m\right)  }p_{11}^{\left(  m\right)  }%
+\eta\left(  1-\eta\right)  p_{10}^{\left(  m\right)  }p_{20}^{\left(
m\right)  }+2\eta p_{00}^{\left(  m\right)  }p_{11}^{\left(  m\right)  }$, respectively.

Then we calculate the (unnormalized) density matrix after the $\left(
m+1\right)  $th ENC,
\end{subequations}
\begin{align}
\tilde{\rho}_{x_{L},y_{R}}^{\left(  m+1\right)  }  &  =\tilde{p}_{00}^{\left(
m+1\right)  }\pi_{00}+\tilde{p}_{10}^{\left(  m+1\right)  }\pi_{10}^{\left(
m+1\right)  }\nonumber\\
&  +\tilde{p}_{11}^{\left(  m+1\right)  }\pi_{11}+\tilde{p}_{20}^{\left(
m+1\right)  }\pi_{20}+\cdots,
\end{align}
with the probability coefficients%
\begin{subequations}
\begin{align}
\tilde{p}_{00}^{\left(  m+1\right)  }  &  \approx\eta p_{10}^{\left(
m\right)  }\left(  p_{00}^{\left(  m\right)  }+\frac{\left(  1-\eta\right)
}{2}p_{10}^{\left(  m\right)  }\right) \\
\tilde{p}_{10}^{\left(  m+1\right)  }  &  \approx\frac{\eta}{2}p_{10}^{\left(
m\right)  }p_{10}^{\left(  m\right)  }\\
\tilde{p}_{11}^{\left(  m+1\right)  }  &  \approx\eta p_{10}^{\left(
m\right)  }p_{11}^{\left(  m+1\right)  }\\
\tilde{p}_{20}^{\left(  m+1\right)  }  &  \approx\frac{\eta}{2}p_{10}^{\left(
m\right)  }p_{20}^{\left(  m+1\right)  }.
\end{align}
to the leading order with respect to the perturbations of $p_{11}^{\left(
m\right)  }$ and $p_{20}^{\left(  m\right)  }$.

Finally, we divide these probabilities by $\tilde{p}_{10}^{\left(  m+1\right)
}$, and obtain\qquad%
\end{subequations}
\begin{subequations}
\begin{align}
\frac{\tilde{p}_{00}^{\left(  m+1\right)  }}{\tilde{p}_{10}^{\left(
m+1\right)  }}  &  \approx2\left(  \frac{p_{00}^{\left(  m\right)  }}%
{p_{10}^{\left(  m\right)  }}+\frac{\left(  1-\eta\right)  }{2}\right)
>2\frac{p_{00}^{\left(  m\right)  }}{p_{10}^{\left(  m\right)  }}\\
\frac{\tilde{p}_{11}^{\left(  m+1\right)  }}{\tilde{p}_{10}^{\left(
m+1\right)  }}  &  \approx2\frac{p_{11}^{\left(  m\right)  }}{p_{10}^{\left(
m\right)  }}\\
\frac{\tilde{p}_{20}^{\left(  m+1\right)  }}{\tilde{p}_{10}^{\left(
m+1\right)  }}  &  \approx\frac{p_{20}^{\left(  m\right)  }}{p_{10}^{\left(
m\right)  }}.
\end{align}

Since the normalization does not change the relative ratio between the
probabilities, the above purterbative estimate tells us that the fractions for
both the vacuum state ($\pi_{00}$) and the multi-excitation state ($\pi_{11}$)
are at least doubled, relative to the logical state $\pi_{10}^{\left(
m+1\right)  }$, after each ENC.

In terms of the notation corresponding to Eq.(\ref{rou_m}) ($p_{vac}=p_{00}$,
$p_{\mathrm{\emph{logic}}}=p_{10}$, and $p_{multi}=p_{11}+p_{20}+\cdots$), we
have%
\end{subequations}
\begin{subequations}
\begin{align}
\frac{p_{vac}^{\left(  m+1\right)  }}{p_{\mathrm{\emph{logic}}}^{\left(
m+1\right)  }}  &  =\frac{\tilde{p}_{00}^{\left(  m+1\right)  }}{\tilde
{p}_{10}^{\left(  m+1\right)  }}>2\frac{p_{vac}^{\left(  m\right)  }%
}{p_{\mathrm{\emph{logic}}}^{\left(  m\right)  }}\\
\frac{\tilde{p}_{multi}^{\left(  m+1\right)  }}{\tilde{p}%
_{\mathrm{\emph{logic}}}^{\left(  m+1\right)  }}  &  \approx2\frac
{p_{multi}^{\left(  m\right)  }}{p_{\mathrm{\emph{logic}}}^{\left(  m\right)
}}%
\end{align}
The ratio of non-logical states to logical states is at least doubling with
distance. As discussed in the main text, it is these unstable non-logical
states that leads to the super-polynomial scaling for the DLCZ protocol.

\section{Non-logical states for the new scheme}

For the new scheme, we can similiarly decompose the density matrix (following
Eq.(\ref{rou_m}))
\end{subequations}
\begin{align}
\rho &  =p_{00}\pi_{00}+p_{10}\pi_{10}+p_{11}\pi_{11}+p_{20}\pi_{20}%
\nonumber\\
&  +p_{21}\pi_{21}+p_{30}\pi_{30}\cdots
\end{align}
where $p_{ij}\pi_{ij}$ is the projected density matrix to the subspace spanned
the Fock states with $i$ (or $j$) photons in $a$ cell-pair and $j$ (or $i$)
photons in $b$ cell-pair, with probability $p_{ij}$ and normalized density
matrix $\pi_{ij}$. For the new scheme, the vacuum type of states consists of
$\pi_{00}$ and $\pi_{10}$, the logical type of state is $\pi_{11}$, and the
rest belong to the multi-excitation type.

We use a perturbative approach, by assuming $p_{00}\pi_{00}$, $p_{10}\pi_{10}%
$, and $p_{11}\pi_{11}$ are the dominant terms, and the rest terms are
perturbations of order $p_{c}$ (terms not listed are of order $p_{c}^{2}$). We
eliminate those irrelevant perturbation terms (e.g. $p_{30}\pi_{30}$), because
after one level of ENC, they are suppressed to $O\left(  p_{c}^{2}\right)  $.

Suppose the entangled state after the $m$th ENC is%
\begin{align}
\rho_{a,b}^{\left(  m\right)  }  &  =p_{00}^{\left(  m\right)  }\pi
_{00}+p_{10}^{\left(  m\right)  }\pi_{10}+p_{11}^{\left(  m\right)  }\pi
_{11}^{\left(  m\right)  }+p_{20,\parallel}^{\left(  m\right)  }%
\pi_{20,\parallel}\nonumber\\
&  +p_{20,\perp}^{\left(  m\right)  }\pi_{20,\perp}+p_{21,\parallel}^{\left(
m\right)  }\pi_{21,\parallel}+p_{21,\perp}^{\left(  m\right)  }\pi_{21,\perp
}+\cdots
\end{align}
Notice that we need to distinguish two possible types of states for $\pi_{20}$
because they behave differently during ENC. The first type of states (denoted
as $\pi_{20,\parallel}$) has both photons stored in the same cell, and after
retrieval the photons will have the same polarization, follow the same path
way, and trigger the photon detector(s) on the same side of the PBS (Fig.
2(b)). The second type of states (denoted as $\pi_{20,\perp}$) has two photons
stored in different cells, and after the retrieval the photons will have
orthogonal polarization, split at the PBS, and trigger photon detectors on
both sides of the PBS (Fig. 2(b)). Thus, the second type of states are more
likely to give the correct click pattern and thus propagate the error to the
next level of ENC. Similarly, we introduce $\pi_{21,\parallel}$ and
$\pi_{21,\perp}$.

In the rest of the discussion, we still follow the convention from the
previous subsection that $\pi_{ij}=\pi_{ji}$ and we will add a superscript $m$
to $\pi_{ij}$ only when we want to keep track of the coherence for that
specific term.

We now connect two such entangled states $\rho_{a_{L},b_{C}}^{\left(
m\right)  }$ and $\rho_{a_{C},b_{R}}^{\left(  m\right)  }$ for the $\left(
m+1\right)  $th ENC%
\begin{equation}
\tilde{\rho}_{a_{L},b_{R}}^{\left(  m+1\right)  }=\mathcal{E}_{ENC}\left[
\rho_{a_{L},b_{C}}^{\left(  m\right)  },\rho_{a_{C},b_{R}}^{\left(  m\right)
}\right]
\end{equation}
where $\tilde{\rho}_{a_{L},b_{R}}^{\left(  m+1\right)  }$ is the unnormalized
density matrix for the entangled state after the $\left(  m+1\right)  $th ENC.

Now we calculate $\mathcal{E}_{ENC}^{sym}\left[  \pi_{\alpha},\pi_{\beta
}\right]  \equiv\frac{1}{2}\mathcal{E}_{ENC}\left[  \pi_{\alpha},\pi_{\beta
}\right]  +\frac{1}{2}\mathcal{E}_{ENC}\left[  \pi_{\beta},\pi_{\alpha
}\right]  $ for $\pi_{\alpha},\pi_{\beta}\in\left\{  \pi_{00},\pi_{10}%
,\pi_{11}^{\left(  m\right)  },\pi_{20,\parallel\text{(or }\perp\text{)}}%
,\pi_{21,\parallel\text{(or }\perp\text{)}},\cdots\right\}  $. For example,%
\begin{subequations}
\begin{align}
\mathcal{E}_{ENC}^{sym}\left[  \pi_{11}^{\left(  m\right)  },\pi_{11}^{\left(
m\right)  }\right]   &  =\frac{\eta^{2}}{2}\pi_{11}^{\left(  m+1\right)
\prime}\\
\mathcal{E}_{ENC}^{sym}\left[  \pi_{11}^{\left(  m\right)  },\pi_{10}\right]
&  =\frac{\eta^{2}}{4}\pi_{10}\\
\mathcal{E}_{ENC}^{sym}\left[  \pi_{11}^{\left(  m\right)  },\pi_{00}\right]
&  =0\\
\mathcal{E}_{ENC}^{sym}\left[  \pi_{11}^{\left(  m\right)  },\pi_{20,\delta
}\right]   &  =\eta^{2}\left(  1-\eta\right)  \pi_{10}\\
\mathcal{E}_{ENC}^{sym}\left[  \pi_{11}^{\left(  m\right)  },\pi_{21,\delta
}\right]   &  =\frac{\eta^{2}}{4}\pi_{21,\delta}+\frac{\eta^{2}\left(
1-\eta\right)  }{2}\pi_{11}\\
\mathcal{E}_{ENC}^{sym}\left[  \pi_{10},\pi_{10}\right]   &  =\frac{\eta^{2}%
}{8}\pi_{00}\\
\mathcal{E}_{ENC}^{sym}\left[  \pi_{10},\pi_{00}\right]   &  =0\\
\mathcal{E}_{ENC}^{sym}\left[  \pi_{10},\pi_{21,\parallel}\right]   &
=\frac{\eta^{2}\left(  1-\eta\right)  }{4}\pi_{10}+\frac{\eta^{2}}{8}%
\pi_{20,\parallel}\\
\mathcal{E}_{ENC}^{sym}\left[  \pi_{10},\pi_{21,\perp}\right]   &  =\frac
{\eta^{2}}{4}\pi_{1,1}+\frac{\eta^{2}\left(  1-\eta\right)  }{4}\pi_{10}%
+\frac{\eta^{2}}{8}\pi_{20,\perp}%
\end{align}%
\begin{align}
\mathcal{E}_{ENC}^{sym}\left[  \pi_{00},\pi_{21,\parallel}\right]   &  =0\\
\mathcal{E}_{ENC}^{sym}\left[  \pi_{00},\pi_{21,\perp}\right]   &  =\frac
{\eta^{2}}{2}\pi_{10}%
\end{align}
The logical type of state after the $\left(  m+1\right)  $th ENC, $\pi
_{11}^{\left(  m+1\right)  }$, is the average of the states $\pi_{11}^{\left(
m+1\right)  \prime}$ and $\pi_{11}$, with relative weights $\frac{\eta^{2}}%
{2}p_{11}^{\left(  m\right)  }p_{11}^{\left(  m\right)  }$ and $\eta
^{2}\left(  1-\eta\right)  p_{11}^{\left(  m\right)  }\sum_{\delta
=\parallel,\perp}p_{21,\delta}^{\left(  m\right)  }+\frac{\eta^{2}}{2}%
p_{10}^{\left(  m\right)  }p_{21,\perp}^{\left(  m\right)  }$, respectively.

Then we calculate the (unnormalized) density matrix after the $\left(
m+1\right)  $th ENC,%
\end{subequations}
\begin{align}
\tilde{\rho}_{a_{L},b_{R}}^{\left(  m+1\right)  }  &  =\tilde{p}_{00}^{\left(
m+1\right)  }\pi_{00}+\tilde{p}_{10}^{\left(  m+1\right)  }\pi_{10}+\tilde
{p}_{11}^{\left(  m+1\right)  }\pi_{11}^{\left(  m+1\right)  }\nonumber\\
&  +\tilde{p}_{20,\parallel}^{\left(  m+1\right)  }\pi_{20,\parallel
}+p_{20,\perp}^{\left(  m+1\right)  }\pi_{20,\perp}+p_{21,\parallel}^{\left(
m+1\right)  }\pi_{21,\parallel}\nonumber\\
&  +p_{21,\perp}^{\left(  m+1\right)  }\pi_{21,\perp}+\cdots
\end{align}
with the probability coefficients%
\begin{subequations}
\begin{align}
\tilde{p}_{00}^{\left(  m+1\right)  }  &  \approx\frac{\eta^{2}}{8}%
p_{10}^{\left(  m\right)  }p_{10}^{\left(  m\right)  }\\
\tilde{p}_{10}^{\left(  m+1\right)  }  &  \approx\frac{\eta^{2}}{2}%
p_{11}^{\left(  m\right)  }p_{10}^{\left(  m\right)  }\times\left(  1+O\left(
p_{c}\right)  \right) \\
\tilde{p}_{11}^{\left(  m+1\right)  }  &  \approx\frac{\eta^{2}}{2}\left(
\begin{array}
[c]{c}%
p_{11}^{\left(  m\right)  }p_{11}^{\left(  m\right)  }+p_{10}^{\left(
m\right)  }p_{21,\perp}^{\left(  m\right)  }\\
+2\left(  1-\eta\right)  p_{11}^{\left(  m\right)  }\left(  p_{21,\parallel
}^{\left(  m\right)  }+p_{21,\perp}^{\left(  m\right)  }\right)
\end{array}
\right) \\
\tilde{p}_{20,\delta}^{\left(  m+1\right)  }  &  \approx\frac{\eta^{2}}%
{4}p_{10}^{\left(  m\right)  }p_{21,\delta}^{\left(  m\right)  }\\
\tilde{p}_{21,\delta}^{\left(  m+1\right)  }  &  \approx\frac{\eta^{2}}%
{2}p_{11}^{\left(  m\right)  }p_{21,\delta}^{\left(  m\right)  }%
\end{align}

Finally, we divide these probabilities by $\tilde{p}_{11}^{\left(  m+1\right)
}$, and to the leading order (i.e. the zeroth order of $p_{c}$) we obtain%
\end{subequations}
\begin{align}
\frac{\tilde{p}_{00}^{\left(  m+1\right)  }}{\tilde{p}_{11}^{\left(
m+1\right)  }}  &  \approx\frac{1}{4}\left(  \frac{p_{10}^{\left(  m\right)
}}{p_{11}^{\left(  m\right)  }}\right)  ^{2}\\
\frac{\tilde{p}_{10}^{\left(  m+1\right)  }}{\tilde{p}_{11}^{\left(
m+1\right)  }}  &  \approx\frac{p_{10}^{\left(  m\right)  }}{p_{11}^{\left(
m\right)  }}\\
\frac{\tilde{p}_{20,\delta}^{\left(  m+1\right)  }}{\tilde{p}_{11}^{\left(
m+1\right)  }}  &  \approx\frac{1}{2}\frac{p_{10}^{\left(  m\right)  }}%
{p_{11}^{\left(  m\right)  }}\frac{p_{21,\delta}^{\left(  m\right)  }}%
{p_{11}^{\left(  m\right)  }}\\
\frac{\tilde{p}_{21,\delta}^{\left(  m+1\right)  }}{\tilde{p}_{11}^{\left(
m+1\right)  }}  &  \approx\frac{p_{21,\delta}^{\left(  m\right)  }}%
{p_{11}^{\left(  m\right)  }}%
\end{align}
Furthermore, all these ratios remain constant (to order $p_{c}$), which
justifies the claim that the probabilities for different types of states
remain stable for all higher levels of ENC.

If we further introduce $\tilde{p}_{vac}^{\left(  m+1\right)  }=\tilde{p}%
_{00}^{\left(  m+1\right)  }+\tilde{p}_{10}^{\left(  m+1\right)  }$,
$\tilde{p}_{\text{logic}}^{\left(  m+1\right)  }=\tilde{p}_{11}^{\left(
m+1\right)  }$and $\tilde{p}_{multi}^{\left(  m+1\right)  }=\sum
_{\delta=\parallel,\perp}\tilde{p}_{20,\delta}^{\left(  m+1\right)  }%
+\tilde{p}_{21,\delta}^{\left(  m+1\right)  }$, we have%
\begin{subequations}
\begin{align}
\tilde{p}_{\mathrm{\emph{logic}}}^{\left(  m+1\right)  }  &  \approx\frac
{\eta^{2}}{2}p_{\mathrm{\emph{logic}}}^{\left(  m\right)  }%
p_{\mathrm{\emph{logic}}}^{\left(  m\right)  }\left(  1+p_{err,new}^{\left(
m+1\right)  }+O\left(  p_{c}\right)  \right) \\
\tilde{p}_{vac}^{\left(  m+1\right)  }  &  \approx\frac{\eta^{2}}%
{2}p_{\mathrm{\emph{logic}}}^{\left(  m\right)  }p_{vac}^{\left(  m\right)
}\left(  1+O\left(  p_{c}\right)  \right) \\
\tilde{p}_{multi}^{\left(  m+1\right)  }  &  \approx\frac{\eta^{2}}%
{2}p_{\mathrm{\emph{logic}}}^{\left(  m\right)  }p_{multi}^{\left(  m\right)
}\left(  1+O\left(  p_{c}\right)  \right)
\end{align}
where $p_{err,new}^{\left(  m+1\right)  }$ is the probability for the new
logical error from the multi-excitation states (accompanied by photon loss)
\end{subequations}
\begin{align}
p_{err,new}^{\left(  m+1\right)  }  &  \approx\frac{2\left(  1-\eta\right)
p_{11}^{\left(  m\right)  }\left(  p_{21,\parallel}^{\left(  m\right)
}+p_{21,\perp}^{\left(  m\right)  }\right)  +p_{10}^{\left(  m\right)
}p_{21,\perp}^{\left(  m\right)  }}{p_{11}^{\left(  m\right)  }p_{11}^{\left(
m\right)  }}\nonumber\\
&  \sim\left(  1-\eta\right)  p_{multi}^{\left(  m\right)  }/p_{\text{logic}%
}^{\left(  m\right)  }\nonumber\\
&  \sim\left(  1-\eta\right)  p_{c}.
\end{align}


\begin{thebibliography}{99}                                                                                               %


\bibitem {GisinRMP}N. Gisin, G. G. Ribordy, W. Tittel, et al., Reviews of
Modern Physics \textbf{74}, 145 (2002).

\bibitem {Kwiat98}W. T. Buttler, R. J. Hughes, P. G. Kwiat, et al., Physical
Review Letters \textbf{81}, 3283 (1998).

\bibitem {Zeilinger04}A. Poppe, A. Fedrizzi, R. Ursin, et al., Optics Express
\textbf{12}, 3865 (2004).

\bibitem {Bennett1996}C. H. Bennett, G. Brassard, S. Popescu, et al., PRL
\textbf{76}, 722 (1996); C. H. Bennett, D. P. DiVincenzo, J. A. Smolin, et
al., PRA \textbf{54}, 3824 (1996).

\bibitem {Deutsch1996}D. Deutsch, A. Ekert, R. Jozsa, et al., PRL \textbf{77},
2818 (1996).

\bibitem {Briegel1998}H.-J. Briegel, W. Dur, J. I. Cirac, et al., PRL
\textbf{81}, 5932 (1998); W. Dur, H. J. Briegel, J. I. Cirac, et al., PRA
\textbf{59}, 169 (1999).

\bibitem {Duan2001}L. M. Duan, M. D. Lukin, J. I. Cirac and P. Zoller, Nature
\textbf{414}, 413 (2001).

\bibitem {Cirac1997}J. I. Cirac, P. Zoller, H. J. Kimble, et al., Physical
Review Letters \textbf{78}, 3221 (1997).

\bibitem {Kimble2004}L. M. Duan and H. J. Kimble, PRL \textbf{92}, 127902 (2004).

\bibitem {Blinov2004}B. B. Blinov, D. L. Moehring, L. M. Duan, et al., Nature
\textbf{428}, 153 (2004).

\bibitem {Childress2004}L. Childress, J. M. Taylor, A. S. Sorensen, et al.,
PRA \textbf{72}, 052330 (2005).

\bibitem {Chou2005}C. W. Chou, H. de Riedmatten, D. Felinto, et al., Nature
\textbf{438}, 828 (2005).

\bibitem {Chaneliere2005}T. Chaneliere, D. N. Matsukevich, S. D. Jenkins, et
al., Nature \textbf{438}, 833 (2005).

\bibitem {Eisaman2005}M. D. Eisaman, A. Andre, F. Massou, et al., Nature
\textbf{438}, 837 (2005).

\bibitem {LukinRMP}M. D. Lukin, Rev. Mod. Phys. \textbf{75}, 457 (2003).

\bibitem {Pan2003}J.-W. Pan, S. Gasparoni, R. Ursin, et al., Nature
\textbf{423}, 417 (2003).

\bibitem {Knill2001}E. Knill, R. Laflamme, and G. J. Milburn, Nature
\textbf{409}, 46(2001).

\bibitem {Merzbacher1998}E. Merzbacher, Quantum Mechanics (John Wiley \& Sons,
Inc., 1998).

\bibitem {Eisaman2004}M. D. Eisaman, L. Childress, A. Andre, et al., Physical
Review Letters \textbf{93, }233602 (2004).

\bibitem {Yamamoto2003}T. Yamamoto, M. Koashi, S. K. Ozdemir, et al., Nature
\textbf{421}, 343 (2003).

\bibitem {Thompson2003}J. K. Thompson, Jonathan Simon, Huanqian Loh, Vladan
Vuleti\'{c}, Science \textbf{313}, 74 (2006).

\bibitem {Collins06}O. A. Collins, S. D. Jenkins, A. Kuzmich, and T. A. B.
Kennedy, archive, quant-ph/0610036.

\bibitem {Chen06}Z. B. Chen, B. Zhao, J. Schmiedmayer et al., quant-ph/0609151.
\end{thebibliography}
\end{document}